\documentclass{PoS}
\usepackage[utf8]{inputenc}
\usepackage{amsmath}

\title{Chiral Perturbation Theory at Finite Volume and/or with Twisted Boundary Conditions}

\ShortTitle{ChPT at finite volume and/or twisting}

\author{\speaker{Johan Bijnens} and Johan Relefors
        \\
        Department of Astronomy and Theoretical Physics, Lund University\\
        S\"olvegatan 14A, SE 22362 Lund, Sweden\\
        E-mail: \email{bijnens@thep.lu.se}}

\abstract{In this talk we discuss a number of ChPT calculations relevant for
          lattice QCD. These include the finite volume corrections at two-loop
           order for masses and decay constants. The second part is about
           hadronic  vacuum polarization where we present the two-loop ChPT
           estimate for the disconnected and strange quark contributions.
           We also present the finite volume corrections at two-loop order.
           The final part is the one-loop finite volume with twisted boundary
           conditions contribution to $f_+(q^2)$ and the full $K_{\ell3}$
           amplitude}

\FullConference{34th annual International Symposium on Lattice Field Theory\\
		24-30 July 2016\\
		University of Southampton, UK}

\renewcommand{\logo}
{\raisebox{-5mm}
{\parbox{\textwidth}{\begin{flushright}
LU TP 16-61\\
November 2016\\
\end{flushright}
\includegraphics{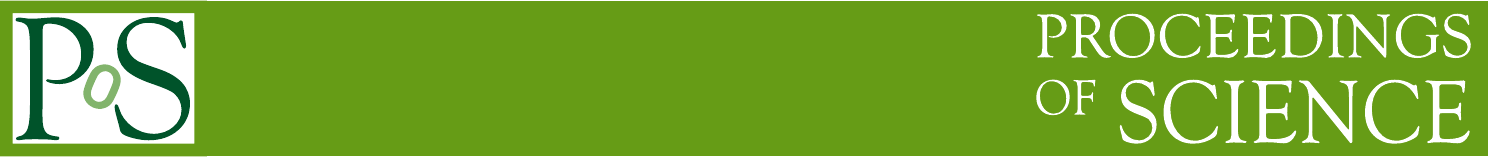}}
}}
\FullConference{{\rm To be published in the proceedings of:\\}
34th annual International Symposium on Lattice Field Theory\\
		24-30 July 2016\\
		University of Southampton, UK}

\begin{document}

\section{Introduction}

This talk discusses some recent applications of Chiral Perturbation Theory
(ChPT) relevant for lattice QCD.
In Sect.~\ref{FVmassesdecay} we discuss the two-loop finite
volume corrections to masses and decay constants for a large number of cases.
The ChPT framework \textsc{CHIRON} that includes most of the results
described in this talk is presented in Sect.~\ref{secchiron}.
The main part is devoted to vector two-point functions where we discuss
results for the disconnected and strange quark contributions and finite volume
corrections at two-loop order including twist effects.
The final section discusses recent work on $K_{\ell3}$ form-factors at one-loop
order in the twisted, staggered and partially quenched case.

The calculations here assume an infinite time extent and are done in the
$p$-regime, i.e. $m_\pi L$ not too small and extra zero mode contributions of the pseudo-scalars are not included. The conventions used are Minkowskian.

\section{Finite volume: masses and decay constants}
\label{FVmassesdecay}

\begin{figure}[t]
\begin{minipage}{0.49\textwidth}
\includegraphics[width=\textwidth]{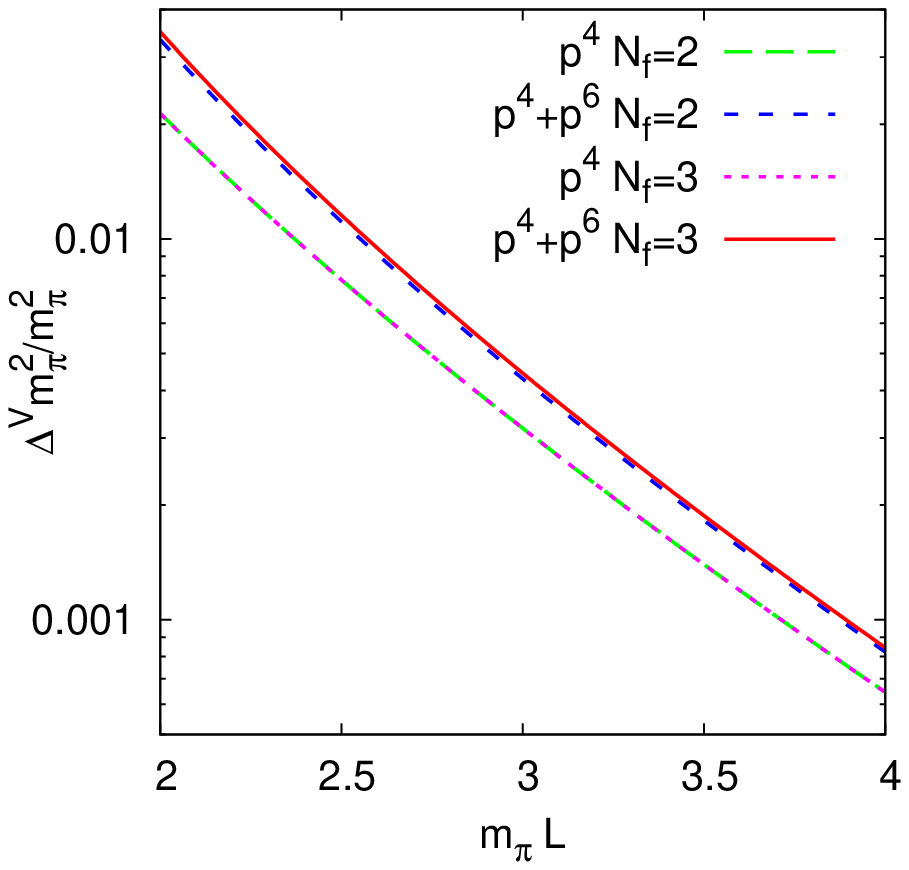}
\vskip-5mm
\centerline{(a)}
\end{minipage}
\begin{minipage}{0.49\textwidth}
\includegraphics[width=\textwidth]{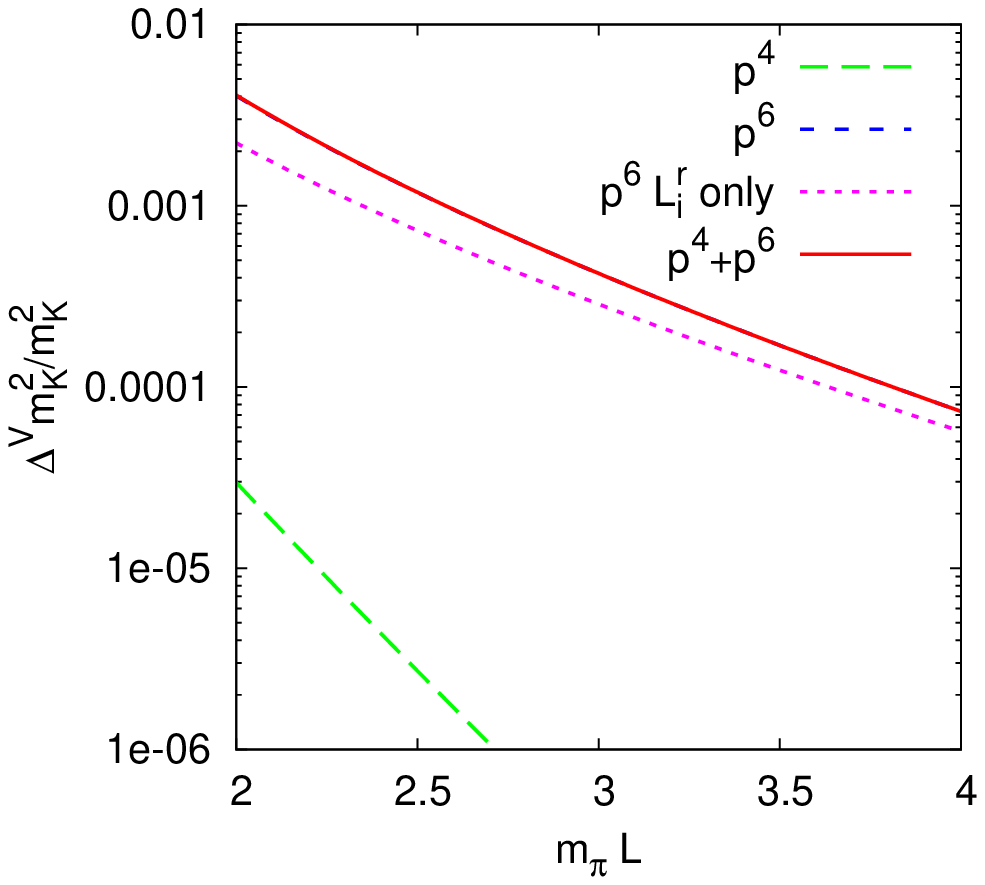}
\vskip-5mm
\centerline{(b)}
\end{minipage}
\caption{\label{figmass} The relative finite volume corrections for the pion
and kaon mass squared.}
\end{figure}

The first applications of ChPT to finite volume (FV) were done
in \cite{Gasser:1986vb}.
The first two-loop FV calculations were the mass in the
two-flavour case \cite{Colangelo:2006mp} and the vacuum-expectation-value
for 3-flavours \cite{Bijnens:2006ve}. Further two-loop progress
was delayed until the two-loop sunset-integrals were worked out for the
general mass case \cite{Bijnens:2013doa}. With these, the FV
corrections at two-loop order were done for the two- and three-flavour case
\cite{Bijnens:2014dea}. As an example of the type of corrections
that were obtained the relative finite volume corrections for the pion mass is
shown in Fig.~\ref{figmass}(a).
The $p^6$-correction is of moderate size and the two- and
three-flavour result are essentially identical. This is as expected
since for the
finite volume corrections the pion-loops are the dominant part
by far. For the kaon mass, Fig.~\ref{figmass}(b), the $p^6$ corrections
is much larger than the one-loop result. The final correction is of
reasonable size. The reason is that for the kaon mass at one-loop there is no
pion contribution. For the decay constant
the same agreement between two- and three-flavour ChPT exists for the
pion and for the kaon the one-loop finite volume correction
to the decay-constant is of normal size since it includes a pion
contribution. The figures can be found in \cite{Bijnens:2014dea}.

With the full result for the sunset-integrals at finite volume a number of
existing calculations at infinite volume at two-loop order could also be
extended to finite volume. The partially quenched ChPT three-flavour results
for masses and decay-constants
of \cite{Bijnens:2004hk,Bijnens:2005ae}
were recalculated and extended to finite volume in
\cite{Bijnens:2015xba}. The two-loop results for masses, decay-constants
and vacuum-expectation-values for QCD-like theories \cite{Bijnens:2009qm}
were extended to the partially quenched case and to finite volume in
\cite{Bijnens:2015dra}. Plots for a number of relevant cases can be found
in those papers.

\section{Chiron}
\label{secchiron}

The main purpose of doing partially quenched and finite volume two-loop ChPT
calculations is that they can be used by the lattice QCD community.
However, the expressions are normally very long. The numerical programs
have been available as a rule from the authors but a more general
numerical framework seemed useful. All results for masses and decay constants
at two-loop order referred to above are available already in the framework
\textsc{CHIRON} \cite{Bijnens:2014gsa}, written in \textsc{C++}.
The library provides a number number of classes for dealing with input
parameters and the ChPT low-energy-constants (LECs). It also contains a large number
of examples.

\section{The vector two-point function and HVP}
\label{HVP}

The lowest-order hadronic-vacuum polarization (HVP)
contribution to the muon anomalous magnetic
moment was discussed a lot at this conference, an overview can be found in
the plenary talk by H.~Wittig ~\cite{Wittig}.
The underlying problem is that one needs values of the two-point function
of electromagnetic currents at low values of  $q^2$ of order $m_\mu^2$ and
below at rather high precision. See e.g. the plot in~\cite{Aubin:2015rzx}.
Two issues are of importance: the size of the more difficult to
calculate disconnected part and the size of the finite-volume corrections.
ChPT can help with both.

\subsection{Disconnected and strange quark contributions}

\begin{figure}[t]
\centerline{\includegraphics[width=8cm]{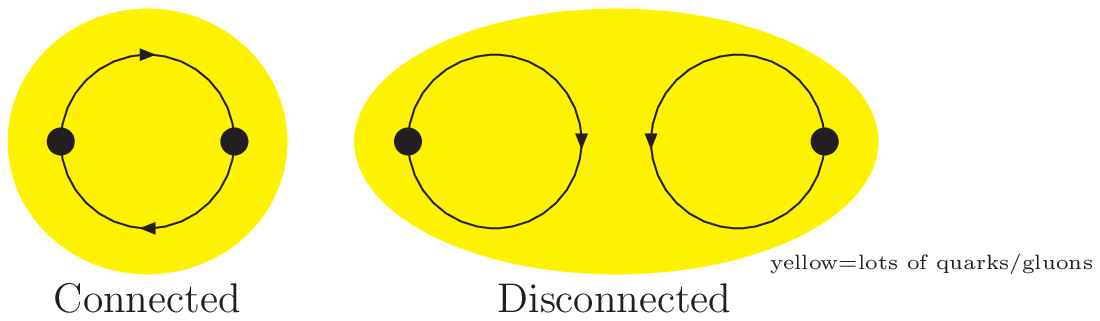}}
\caption{\label{figdisconnected} A schematic drawing of disconnected
and connected contributions. The dots are insertions of the currents
and the lines are the valence quarks. The background indicates the sea-quarks
and gluons.}
\end{figure}

The underlying object is the two-point function of vector currents:
\begin{equation}
  \Pi^{\mu\nu}_{ab}(q) \equiv i\int d^4x e^{iq\cdot x}
\big<T(j^\mu_a(x)j_a^{\nu\dagger}(0))\big>\,.
\end{equation}
For HVP we need it with the electromagnetic current with two- or
three- flavours of light quarks, but we also define a number of simpler currents
\begin{align}
j_{\pi^+}^\mu = & \bar d\gamma^\mu u\,, &
j_{u}^\mu = & \bar u\gamma^\mu u\,,     &
j_{d}^\mu = & \bar d\gamma^\mu d\,, &
j_{s}^\mu = & \bar s\gamma^\mu s\,, &
j_{e}^\mu = &\frac{2}{3}\bar u\gamma^\mu u-\frac{1}{3}\bar d\gamma^\mu d
\left(-\frac{1}{3}\bar s\gamma^\mu s\right)\,.
\end{align}
In lattice QCD the two-point functions have two types of contributions:
connected or disconnected shown schematically in Fig.~\ref{figdisconnected}.
The size of the different contributions has been discussed
at one-loop order in ChPT in \cite{DellaMorte:2010aq}.
In this talk and \cite{Bijnens:2016ndo} we have extended their results
to two-loop order and generalized it in a number of other ways.
References to other papers can be found in \cite{Wittig,Bijnens:2016ndo}.

We will use two main observations in order to estimate the size of disconnected
to connected contributions. The first is that the singlet vector current
does not couple to mesons until rather high order in ChPT.
The coupling starts at $p^4$ via the Wess-Zumino-Witten term but this
contributes only at order $p^{10}$, the normal coupling starts at order
$p^6$ and thus only contributes at order $p^8$. As a consequence, only direct
counter-term contributions are present for the singlet current at order $p^4$
and $p^6$. These can change the ratios discussed below.
However, a large part of higher
loop corrections will have the same ratio.

For the two-flavour case we have that $\Pi^{\mu\nu}_{\pi^+\pi^+}$
is fully connected, $\Pi^{\mu\nu}_{ud}$ only disconnected and
$\Pi^{\mu\nu}_{uu}$ is the sum of the two. Since the singlet current does
not couple we also have that $\Pi^{\mu\nu}_{(u+d)u}=0$ for the contributions
mentioned above. The disconnected part is thus $-1/2$ time the connected part
for those contributions. The relation
$\Pi^{\mu\nu}_{ee} = \frac{5}{9}\Pi^{\mu\nu}_{\pi^+\pi^+}+\frac{1}{9}\Pi^{\mu\nu}_{ud}$
implies that for $\Pi^{\mu\nu}_{ee}$ the
disconnected part is $-1/10$ of the connected part for those contributions.

\begin{figure}[t]
\begin{minipage}{0.49\textwidth}
\includegraphics[width=\textwidth]{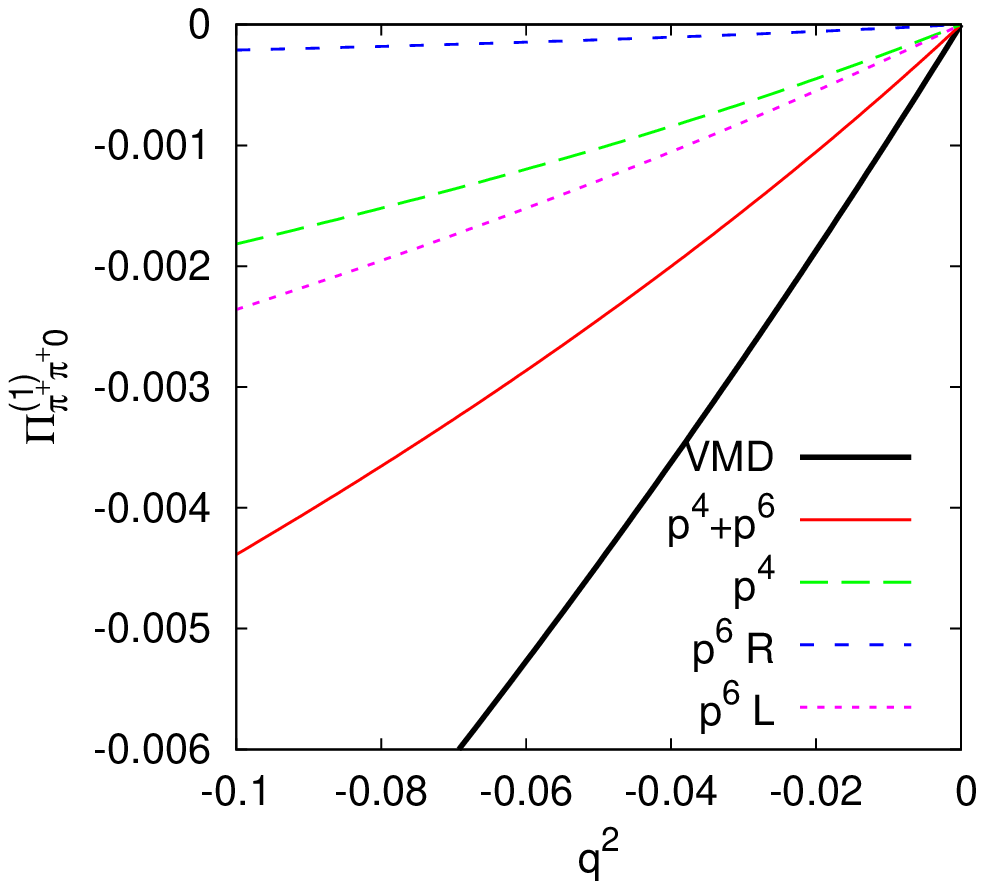}
\vskip-5mm
\centerline{(a)}
\end{minipage}
\begin{minipage}{0.49\textwidth}
\includegraphics[width=\textwidth]{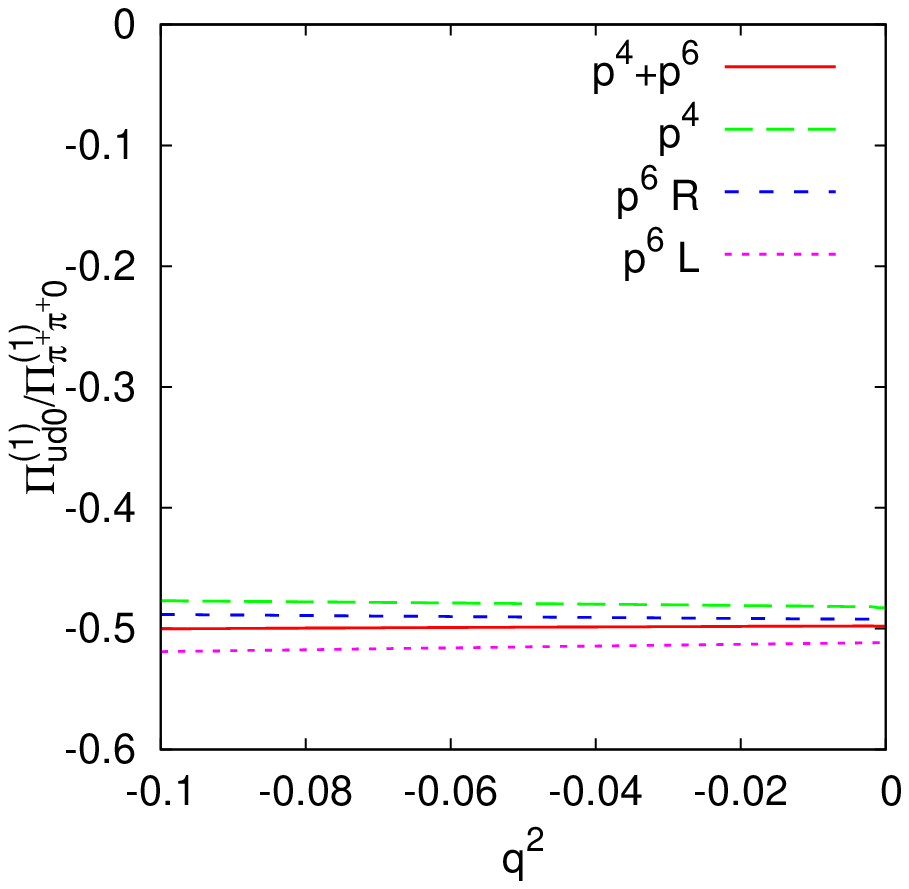}
\vskip-5mm
\centerline{(b)}
\end{minipage}
\caption{\label{figPiconnected} (a) The different contributions
to $\hat\Pi_{\pi^+\pi^+}(q^2)$, i.e. the connected part.
(b) The Ratio of the disconnected to connected part for the different
contributions but without adding the VMD part to the connected.}
\end{figure}

There are new LECs that have couplings to the singlet current. Using a VMD
like estimate we find that the LEC contributions only contribute to the
connected part. These will thus lower the ratio even further.
In infinite volume the two-point functions are given by
$\Pi^{\mu\nu}_{ab} = \left(q^\mu q^\nu - q^2 g^{\mu\nu}\right)
\Pi^{(1)}_{ab}$
and for HVP $\hat\Pi=\Pi^{(1)}(q^2)-\Pi^{(1)}(0)$ is relevant.
This is what is plotted in Fig.~\ref{figPiconnected}.
For the VMD estimate we use
$\Pi^{(1)}_{\pi^+\pi^+} = {4 F_\pi^2}/(M_V^2-q^2)$. The loop contributions
as well as the VMD contributions for the connected part are shown
in Fig.~\ref{figPiconnected}(a).
The ratio of disconnected to connected is shown in Fig.~\ref{figPiconnected}(b).
The pure pion part is exactly $-1/2$ and kaon and eta loops
give only small corrections. Adding the VMD contributions lowers
the ratio significantly as shown in \cite{Bijnens:2016ndo}.

The same method can be used to calculate the strange quark contribution.
Here we find a very strong cancellation between $p^4$ and $p^6$ contributions.
We end up estimating the main part from $\phi$-exchange. More numerical results
and plots can be found in \cite{Bijnens:2016ndo}.
The partially quenched case has been done as
well \cite{Bijnens:2016soon,thesisjohan}, allowing to study the
disconnected strange quark contributions.

\subsection{Twisting}

\begin{figure}[t]
\begin{minipage}{0.49\textwidth}
\includegraphics[width=\textwidth]{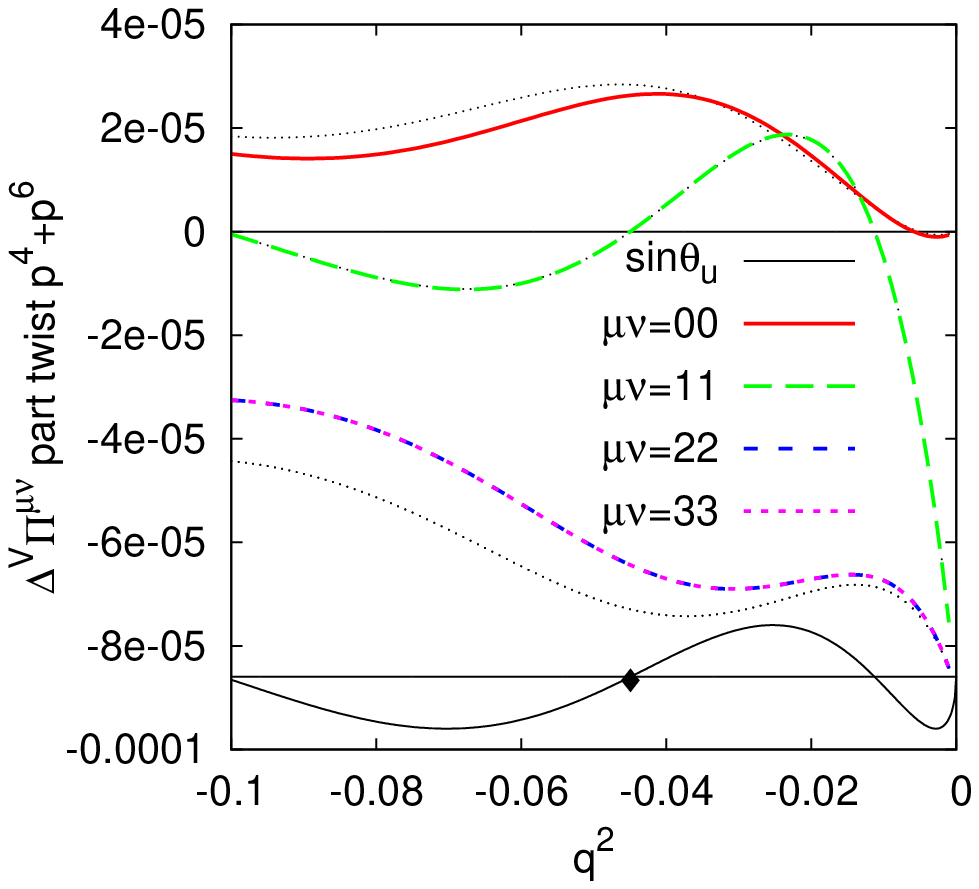}
\centerline{(a)}
\end{minipage}
\begin{minipage}{0.49\textwidth}
\includegraphics[width=\textwidth]{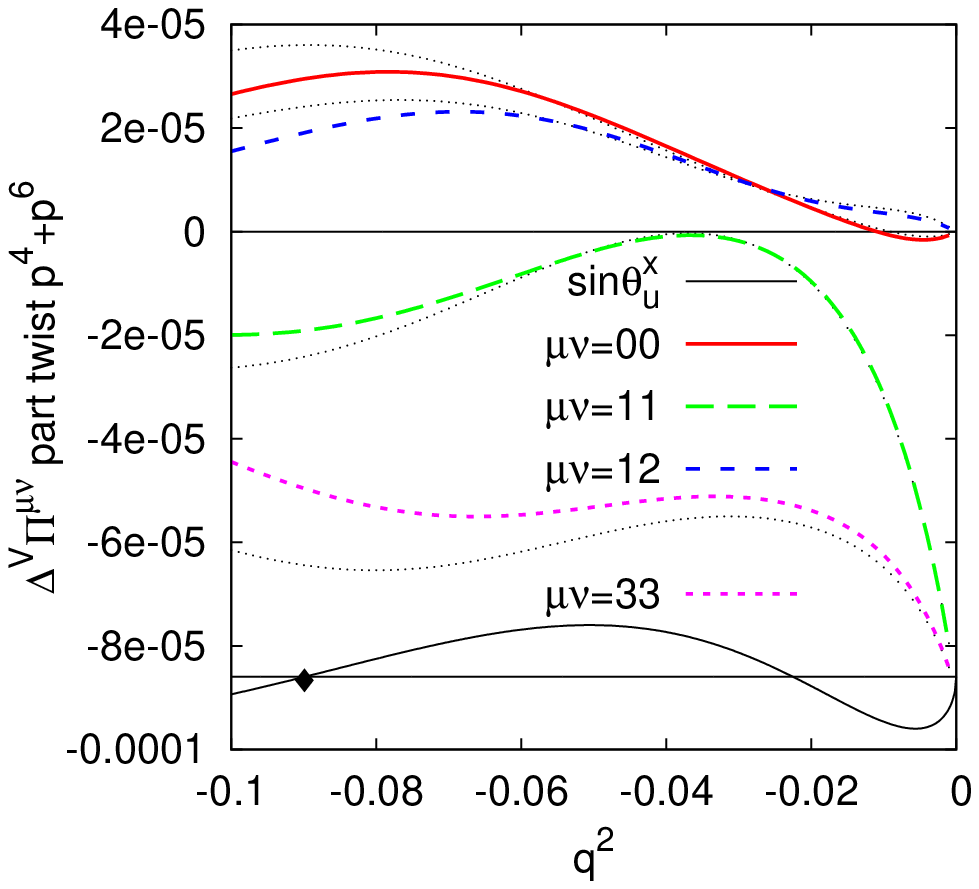}
\centerline{(b)}
\end{minipage}
\caption{\label{figHVP} 
The finite volume contributions with twisted boundary conditions
for several components of $\Pi^{\mu\nu}_{\pi^+\pi^+}(q)$. Shown are the
partially twisted up quark results to two-loop order. The one-loop result
is always shown as the nearest thin line.
The plot is for $m_\pi L=4$.
(a) $q=\left(0,\sqrt{-q^2},0,0\right)$
(b) $q=\left(0,\frac{\sqrt{-q^2}}{\sqrt{2}},\frac{\sqrt{-q^2}}{\sqrt{2}},0\right)$.}
\end{figure}

In a finite volume with periodic boundary conditions only discrete momenta
$p^i = 2\pi n^i/L$ with $n^i$ integer are acceptable.
Twisted boundary conditions for certain quarks
$q(x^i+L) = e^{i\theta^i_q}q(x^i)$
allow instead for $p^i = \theta^i/L + 2\pi n^i/L$.
Using this a continuous momentum space can be mapped out.
ChPT for these boundary conditions was introduced in
\cite{Sachrajda:2004mi}. However, this reduces the cubic symmetry imposed
by the cubic box even further. In general one should thus remember to take this
into account, most quantities can depend on all components of the spatial
momentum and there are many more form-factors in general. In addition, charge
conjugation involves a change of momentum\footnote{Note that because
adding a constant background field is equivalent to imposing twisted boundary
conditions, what is called momentum is somewhat ambiguous. We always use
the above $p^i = \theta^i/L + 2\pi n^i/L$.}.
In particular the two-point function
$\Pi^{\mu\nu}_{a}(q) \equiv i\int d^4x e^{iq\cdot x} 
\big<T(j^\mu_a(x)j_a^{\nu\dagger}(0))\big>$ with
 $j_{\pi^+}^\mu = \bar d\gamma^\mu u$ satisfies
$
\partial_\mu\big<T(j_{\pi^+}^\mu(x)j_{\pi^+}^{\nu\dagger}(0))\big>=
\delta^{(4)}(x)\big<\bar d\gamma^\nu d-\bar u\gamma^\nu u\big>\,.
$
This leads to the Ward-Takahashi identity
$
q_\mu\Pi_{\pi^+}^{\mu\nu}= \big<\bar u\gamma^\mu u-\bar d\gamma^\mu d\big>\,.
$
Because of the twisted boundary conditions the r.h.s. can be nonzero.
Our one- \cite{Bijnens:2014yya}  and two-loop
results \cite{Bijnens:2016soon,thesisjohan}
for finite volume with twisted boundary conditions
satisfy the Ward identity. Related discussions at one-loop can be found in 
\cite{Aubin:2013daa}. Our main new result here is that the two-loop
corrections for the finite volume corrections are of reasonable size.
The is shown for two different ways of choosing the twist angle
in Fig.~\ref{figHVP}.
The plots are shown for the partially twisted case, i.e. only the valence up
quark has a twisted boundary condition. There is in fact very little numerical
difference between the fully twisted, also the sea up-quark twisted, and the
partially twisted case. As one can see, the finite volume corrections for the
two different twists shown in (a)  and (b) in Fig.~\ref{figHVP}, at the same
value of $q^2$, are different. This allows for testing the finite volume
calculation using the same underlying lattice configurations.
Another example is shown in \cite{Bijnens:2016soon,thesisjohan}.

\section{$K_{\ell3}$}
\label{Kl3}

The decay $K\to\pi\ell\nu$ is one of the major sources for determining the
CKM element $|V_{us}|$. To do this requires knowledge to high precision of
the form-factor $f_+$ at $q^2=0$. Lattice QCD is now the most precise
way to determine this and one main remaining error is the finite volume
correction. As in the previous section, partial twisting is often employed,
here such that the form-factor at $q^2=0$ can be calculated directly.
The work discussed in this section is in \cite{Bernard:2016soon,thesisjohan}.

Earlier ChPT work on $K_{\ell3}$ and related form-factors is the original
one-loop work \cite{Gasser:1984ux}, the two-loop work
without \cite{Bijnens:2003uy} and with \cite{Bijnens:2007xa} isospin breaking
and the one-loop partially quenched staggered calculation
\cite{Bernard:2013eya}. Earlier work on finite volume corrections
is \cite{Ghorbani:2013yh}.

As in the previous section, finite volume and twisting lead to more form-factors
and a different Ward-Takahashi identity. With the scalar and vector
form-factors of $K$ to $\pi$ transitions defined as
\begin{align*}
\label{defKl3}
\langle\pi^-(p^\prime)|\bar s\gamma_\mu u(0)|K^0(p)\rangle
= & f_+ (p_\mu+p^\prime_\mu)
 + f_-q_\mu + h_\mu\,,
&
\langle\pi^-(p^\prime)|(m_s-m_u)\bar s u(0)|K^0(p)\rangle
= &\rho\,,
\end{align*}
and using $q=p-p^\prime$ the Ward identity becomes
$(p^2-p^{\prime2})f_+ + q^2 f_- + q^\mu h_\mu = \rho$.
Note the presence of extra terms compared to the infinite volume case
in both equations. The split between $f_+,f_-$ and $h_\mu$ is not unique
but is useful in presenting results.

In \cite{Bernard:2016soon,thesisjohan} we have calculated the form-factors
$f_+,f_-,h_\mu$ and $\rho$ in partially quenched, twisted and staggered ChPT.
Our result satisfy the Ward identity analytically and numerically.
Formulas, plots and more numerical results can be found in
\cite{Bernard:2016soon,thesisjohan}. Here we show
numerical results for the finite volume corrections corresponding to 2 MILC
lattices. The relevant parameters can be found in the top Tab.~\ref{tabresults}.
The numerical results are shown for all terms in the Ward identity but such that
the $f_+$ term is normalized to one. 
In Tab.~\ref{tabresults} we show results for three different cases:
staggered and partially twisted with an up twist angle the same in the three
spatial direction, staggered and ``unstaggered'' with a partially twisted
up-quark in one spatial direction.
Note the different finite volume corrections for the different twist cases.
These can again be used to test the finite volume corrections using the
same underlying lattice configurations.
\begin{table}
\begin{center}
\begin{tabular}{ccrrrr}
\hline
a(fm) & $m_l/m_s$ & L(fm) & $m_\pi$(MeV) & $m_K$(MeV) & $m_\pi L$\\ 
\hline
0.06 & 0.2   & 2.8 & 319 & 547 & 4.5\\
     & 0.035 & 5.5 & 134 & 491 & 3.7\\
\hline
 $m_\pi$ & $m_\pi L$ & ``mass''~ & ``$f_+$''~~ & ``$h_\mu$''~~ & ``$\rho$''~~~\\ 
\hline
 319 & 4.5 & $ 0.00052$&   $0.00037$&$-0.00081$& $ 0.00008$ \\
 134 & 3.7 & $-0.00016$&   $0.00045$& $0.00013$& $ 0.00043$ \\
\hline
 319 & 4.5 & $-0.00026$&   $0.00013$&$-0.00012$& $-0.00025$ \\
 134 & 3.7 & $-0.00005$&  $-0.00058$& $0.00001$& $-0.00062$ \\
\hline
 319 & 4.5 & $-0.00031$&   $0.00015$&$-0.00011$& $-0.00027$ \\
 134 & 3.7 & $-0.00007$&  $-0.00064$& $0.00001$& $-0.00070$ \\
\hline
\end{tabular}
\end{center}
\caption{\label{tabresults}
The finite volume results for $K_{\ell3}$.
The top gives the parameters for two MILC lattices.
The bottom gives the numbers for the different parts of the Ward identity
$\left[
 ({\Delta^V m_K^2-\Delta^V m_\pi^2})/({m_K^2-m_\pi^2})\right]+
 \left[\Delta^V f_+(0)\right]
+\left[{q_\mu h^\mu}/({m_K^2-m_\pi^2})\right] = 
\left[{\Delta^V\rho}/({m_K^2-m_\pi^2})\right]
$ for the above two lattices.
The first set is with a spatially symmetric twist angle
$\theta_u=(0,\theta,\theta,\theta)$.
The two other sets are with a twist in one direction only,
$\theta_u=(0,\theta,0,0)$.
The second set is staggered and the last ``unstaggered.''}
\end{table}

\section*{Acknowledgements}
Work supported in part by Swedish Research Council grants
 621-2013-4287 and 2015-04089 and by
the European Research Council (ERC) grant No 668679.
\textsc{FORM} \cite{FORM} was essential.

\end{document}